\documentclass[hyper]{JHEP} 

\usepackage{epsfig}

\newcommand\fverb{\setbox\pippobox=\hbox\bgroup\verb}
\newcommand\fverbdo{\egroup\medskip\noindent%
			\fbox{\unhbox\pippobox}\ }
\newcommand\fverbit{\egroup\item[\fbox{\unhbox\pippobox}]}
\newbox\pippobox
\title{Vacuum States in 
  2D Tachyon Effective Action}
\author{by J. Kluso\v{n}\\
	 Department of Theoretical Physics and Astrophysics\\
                   Faculty of Science, Masaryk University\\
Kotl\'{a}\v{r}sk\'{a} 2, 611 37, Brno\\
Czech Republic\\
	E-mail: \email{klu@physics.muni.cz}}
\preprint{\hepth{0406265}}

\abstract{In this paper we will  study  ground
states of the
 toy model  of  2D closed string tachyon effective action. 
We will firstly construct the classical solutions
of the tachyon effective action that do not induce
backreaction on metric and dilaton. Then 
we will study   the quantum mechanics of the zero mode
of the tachyon field. We will find family of vacuum states that
are labeled with single parameter. 
We will also perform the  quantum mechanical analysis
of the tachyon effective action when we take into account
dynamics of nonzero modes. We will calculate the vacuum expectation
values of components of the stress energy tensor and
dilaton source  and we will argue that  there is
 not  any backreaction on metric and dilaton.}

\def\bra #1{\left<#1\right|}
\def\ket #1{\left|#1\right>}

\def\bb{\mathbf{B}}

\def\ss{\sin \frac{\tau}{\sqrt{2}}}
\def\st{\sinh \frac{\tau}{\sqrt{2}}}
\def\st2{\sinh^2 \frac{\tau}{\sqrt{2}}}
\def\ss2{\sin^2 \frac{\tau}{\sqrt{2}}}

\def\p{\phi}
\def\hS{\hat{S}}
\def\hP{\hat{\Pi}}

\begin{document}
\section{Introduction}\label{first}
To find off-shell description of 
closed string is very difficult task. The
most interesting exception is noncritical
string theory in two dimensions
\footnote{For reviews, see 
\cite{Klebanov:1991qa,Das:1992dm,Jevicki:1993qn,Ginsparg:is,
Polchinski:1994mb,Martinec:2003ka,Alexandrov:2003ut,Nakayama:2004vk}.}.
Here, there is a complete nonperturbative description in
terms of the double scaling limit of matrix quantum
mechanics. For matrix quantum mechanics the space of
eigenvalues of the matrix  provides the space dimensions
on which free string moves. In this case, however,
the string theory is rather trivial: for a bosonic
string the only dynamical degree of freedom is a single
massless scalar which is related to the collective
field-the density of eigenvalues. The eigenvalues
of this theory behave as fermions and the collective
field may be regarded as a bosonisation of
these fermions. In summary, the discovery of
the double scaling limit of $c<1$ matrix models
provided beautiful and exact solution to the bosonic
string theory to $D\leq 2$ spacetime dimensions.  
However one disappointing property of these models
was the fact that bosonic string theory  
did not appear to be well defined non-perturbativelly.
Recently this problem  appears to have been
removed by reinterpretation of these models
as nonperturbatively well defined type 0 strings
\footnote{For extensive list of references, see
excellent recent review  \cite{Nakayama:2004vk}.}
\cite{Takayanagi:2003sm,Douglas:2003up}.

The two dimensional (2D) bosonic theory is defined
in the linear dilaton background (In units,
where $\alpha'=1$)
\footnote{For review, see \cite{Polchinski:rr}.}
\begin{equation}\label{lindili}
g_{\mu\nu}=\eta_{\mu\nu} \ , 
\Phi=2x^1  \ . 
\end{equation} 
It is well known that dilaton is generator
of the string coupling in the sense $g_s\sim
e^{\Phi}$ so that for the solution
(\ref{lindili}) there is a region where
the coupling constant diverges and
string perturbative theory fails. 
This means that such a background does not
define satisfactory string theory. It turns
out however that this can be cured by 
inclusion of the tachyon condensate 
in the form 
\begin{equation}\label{Tlini}
T=\mu e^{2x^1}  \ , 
\end{equation}
where $\mu$ is  ``cosmological
constant''. The claim is that 
(\ref{lindili}) and (\ref{Tlini}) defines
exact string theory background. Indeed, in
this background the string sigma model takes
the form
\begin{equation}\label{liv}
S=\frac{1}{4\pi}\int d^2\sigma
\sqrt{h}\left[h^{ab}\partial_aX^{\mu}
\partial_bX_{\mu}+2\mathcal{R}X^1+\mu
e^{2X^1}\right]\ .
\end{equation}
It can be checked that this action represents
exact conformal field theory and hence defines
consistent string theory. 

It  is certainly important to find 
an action that has background fields (\ref{lindili}),
 (\ref{Tlini}) as its exact solution.
This intuition is based on the fact that
(\ref{liv}) defines exact CFT and hence
 $\beta$ functions that we  associate
with each field in the worldsheet theory
vanish. Then it is believed that these
equations arise from variation of some
spacetime effective action. However to find such
effective actions from the first principles
of string theory is very difficult task.
Another possibility is to  try to guess their form from
the requirement that the background that
defines exact conformal field theory is
solution of the equation of motions that
arises from it. Such  approach was
very useful for the searching of the
tachyon effective actions in case of unstable
D-branes in string theories
\cite{Kutasov:2003er,Niarchos:2004rw,Kluson:2004qy,Smedback:2003ur}
that has marginal tachyon profile as its
exact solution\footnote{We must also stress that
it is not currently clear how such 
tachyon effective actions are related to
these that were calculated from the first
principles of the string theory
\cite{Tseytlin:2000mt}. For more
detailed discussion of this issue, see 
\cite{Fotopoulos:2003yt}.}.
One can then ask the question whether
similar approach could be applied in the
case of closed string tachyon effective
action. In fact, preliminary 
 steps in this direction
were given recently in  \cite{Kluson:2004ns}.
The basic property of the proposed closed string
 tachyon effective action given there
 is that it  has the marginal tachyon
profile $T=\mu e^{\beta_{\mu}x^{\mu}}$ in
the linear dilaton background as its exact
solution. We have also shown that
at two dimensions the situation simplifies
considerably and we were able to find field theory effective
 action for tachyon,
dilaton and metric that has the linear
dilaton background (\ref{lindili}),
(\ref{Tlini}) as its exact solution even
if we take into account the backreaction
of the tachyon on metric and dilaton. 

In this paper we will continue the study
of this  model of  2D tachyon effective
action. Namely, we focus on the ground
states of this action, either from classical
or quantum point of view. 
As we will  demonstrate
in the next section the tachyon effective
action has remarkable property that 
if we begin with the tachyon background
with some cosmological constant $\mu_1$ in
(\ref{Tlini}) then we 
can easily  reach
 the other tachyon background 
(\ref{Tlini}) characterised
with the second cosmological constant
$\mu_2$.  We hope that this is sign
of the  background independence
of the tachyon effective action.
At first sight it seems that this is rather
trivial fact from the point of view
of the tachyon effective field theory. 
On the other hand we mean that such a form
of background independence is not
directly visible in the fundamental formulation
of two dimensional string theory, either
in free fermion description or in the
collective field theory description. 
For that reason we believe that  the 
 study of the background
independence in the tachyon effective action
can be useful for  understanding of the 
fundamental properties of the   microscopic
formulation of 2D
string theory.

Since generally the free fermion formulation
of the two dimensional string theory is
quantum theory and the parameter 
$\mu$ appears in the definition of the   Fermi sea 
 it seems to be natural
to study   
the ground states in the quantum mechanical
formulation of the tachyon effective action.
 We will see that there exists family
of vacuum states  labelled by single
 parameter that is directly related to 
 $\mu$ in (\ref{Tlini}). 
We will also calculate the expectation
values of the stress energy tensor and dilaton
source in these states and confirm that
there is not any backreaction on metric and
dilaton in agreement with classical analysis.
We will also construct an operator that
maps  one vacuum state to another
one. We think  that the existence of this operator 
 confirms the presumption that 
all ground states with different
values of $\mu$ are equivalent.

 We must also stress one important limitation of our
analysis. As is well known from the effective
field theory description of the  unstable
D-branes the validity of the tachyon effective
actions strongly depends on the region in the field
theory space where they are defined. 
Then it is clear  that the
 model of the tachyon effective action that
was proposed in 
\cite{Kluson:2004ns} cannot describe
 the full dynamics of
the closed string tachyon. Rather we should restrict to the
modes that propagate very close to the classical
 solution (\ref{Tlini}). However  we still  believe that
conclusions considering the ground
states of the tachyon  effective action  are more
general and could be helpful in the study
of  general properties of the tachyon  dynamics. 

The structure of this paper is as follows.
In the next section (\ref{second}) we give the
brief review of the tachyon effective action that
was suggested in \cite{Kluson:2004ns}.
In section (\ref{third}) we  study the
quantum mechanics of the zero mode 
of the tachyon field. In section (\ref{fourth})
we extend this analysis to  the full quantum field theory
of the tachyon effective action and we 
calculate the vacuum expectation values of
the stress energy tensor and dilaton source. 
In conclusion (\ref{fifth}) we outline
our results and suggest further extension
of this work.

\section{Toy model of 2D tachyon
effective action}\label{second}
In this section we   review 
basic facts about  the
toy model of 2D closed string tachyon effective action
suggested in   \cite{Kluson:2004ns}.

As it is 
 well known  spacetime fields of two dimensional
bosonic string theory 
consist  dilaton, graviton and tachyon.
The dilaton and graviton are described with
the action
\footnote{Our convention is as follows. We work
with two dimensional metric with signature
$(-,+)$. The spatial coordinate is labelled with
$x^1\equiv \phi $ and time coordinate is $x^0\equiv t$.}
\begin{equation}\label{metdil}
S_{g,\Phi}=-\int d^2x\sqrt{-g}e^{-2\Phi}
\left(16+R+4g^{\mu\nu}\partial_{\mu}\Phi
\partial_{\nu}\Phi\right) \ .
\end{equation}
In \cite{Kluson:2004ns} we have proposed
 the tachyon effective action that
has the marginal tachyon profile $T=\mu e^{2x^1}$
in the linear dilaton background $\Phi=2x^1$ as
its exact solution even if we take into account
the backreaction of the tachyon on the dilaton and
the graviton. This action has the following form
\begin{eqnarray}\label{Ttwo}
S_T=-\int d^2x\frac{\sqrt{-g}}{
(1+e^{-2\Phi}T^2(4-g^{\mu\nu}\partial_{\mu}\Phi
\partial_{\nu}\Phi))}
\times \nonumber \\
\times \sqrt{1+e^{-2\Phi}\left(4T^2+g^{\mu\nu}
\partial_{\mu}T\partial_{\nu}T-2T
g^{\mu\nu}\partial_{\mu}T\partial_{\nu}\Phi
\right)}+  \nonumber \\
+\int d^2x \sqrt{-g}\frac{1
}
{1+\frac{1}{2}T^2e^{-2\Phi}(4-\partial_{\mu}\Phi g^{\mu\nu}
\partial_{\nu}\Phi)} \ .  \nonumber \\  
\end{eqnarray}
The variation of the action $S=S_{g,\Phi}+S_T$
with respect to $g^{\mu\nu}$ gives
the equation of motion for $g_{\mu\nu}$
\begin{equation}\label{gequation}
e^{-2\Phi}\left(G_{\mu\nu}-
2g_{\mu\nu}\nabla^2\Phi+2\nabla_{\mu}
\nabla_{\nu}\Phi+2g_{\mu\nu}
\left(\nabla \Phi\right)^2-8g_{\mu\nu}\right)=
T_{\mu\nu}^T \ 
\end{equation}
and the variation with respect to $\Phi$ gives
\begin{equation}\label{phiequat}
\sqrt{-g}e^{-2\Phi}
\left[32+2R-8g^{\mu\nu}\partial_{\mu}\Phi
\partial_{\nu}\Phi\right]
+16e^{-2\Phi}\partial_{\mu}
\left[\sqrt{-g}g^{\mu\nu}\partial_{\nu}
\Phi\right]=J_{\Phi} \ ,
\end{equation}
where
\begin{equation}
T^T_{\mu\nu}=-\frac{2}{\sqrt{-g}}\frac{\delta S_T}{\delta g^{\mu\nu}} \ ,
J_{\Phi}=\frac{\delta S_T}{\delta \Phi} \ .
\end{equation}
The explicit forms of the stress energy tensor
and dilaton source are 
\begin{eqnarray}\label{Tstress}
T_{\mu\nu}=-g_{\mu\nu}
\frac{1}{
(1+e^{-2\Phi}T^2(4-
\partial_{\mu}\Phi g^{\mu\nu}\partial_{\nu}
\Phi))}
\sqrt{\bb}
+\nonumber \\
+\frac{2e^{-2\Phi}T^2
\partial_{\mu}\Phi\partial_{\nu}\Phi}{
(1+e^{-2\Phi}T^2(4-
\partial_{\mu}\Phi g^{\mu\nu}\partial_{\nu}
\Phi))^2}
\sqrt{\bb}+
\nonumber \\
+\frac{e^{-2\Phi}(\partial_{\mu}T\partial_{\nu}T
-T\left(
\partial_{\mu}T\partial_{\nu}\Phi
+\partial_{\mu}\Phi\partial_{\nu}T
\right))}{
(1+e^{-2\Phi}T^2(4-
\partial_{\mu}\Phi g^{\mu\nu}\partial_{\nu}
\Phi))}
\sqrt{\bb} \nonumber \\
-g_{\mu\nu}\frac{1}
{1+\frac{1}{2}T^2e^{-2\Phi}(4-g^{\mu\nu}
\partial_{\mu}\Phi\partial_{\nu}\Phi)}
-\frac{e^{-2\Phi}T^2\partial_{\mu}
\Phi\partial_{\nu}\Phi}{
(1+T^2e^{-2\Phi}\frac{1}{2}(4-g^{\mu\nu}
\partial_{\mu}\Phi\partial_{\nu}\Phi))^2} \ ,
\nonumber \\
\end{eqnarray}
\begin{eqnarray}\label{dilsor}
J_{\Phi}=-\frac{2\sqrt{-g}e^{-2\Phi}T^2
(4-\partial_{\mu}\Phi g^{\mu\nu}
\partial_{\nu}\Phi)\sqrt{\bb}}
{(1+e^{-2\Phi}T^2(4-
\partial_{\mu}\Phi g^{\mu\nu}\partial_{\nu}
\Phi))^2}+
\nonumber \\
+\partial_{\mu}\left[
\frac{2\sqrt{-g}e^{-2\Phi}T^2g^{\mu\nu}\partial_{\nu}\Phi
\sqrt{\bb}}
{(1+e^{-2\Phi}T^2(4-
\partial_{\mu}\Phi g^{\mu\nu}\partial_{\nu}
\Phi))^2}\right]+
\nonumber \\
+\frac{\sqrt{-g}e^{-2\Phi}(4T^2+g^{\mu\nu}\partial_{\mu}T
\partial_{\nu}T-2Tg^{\mu\nu}\partial_{\mu}T
\partial_{\nu}\Phi)}{
(1+e^{-2\Phi}T^2(4-
\partial_{\mu}\Phi g^{\mu\nu}\partial_{\nu}
\Phi))\sqrt{\bb}}-
\nonumber \\
-\partial_{\mu}\left[\frac{\sqrt{-g}e^{-2\Phi}
Tg^{\mu\nu}\partial_{\nu}T}
{(1+e^{-2\Phi}T^2(4-
\partial_{\mu}\Phi g^{\mu\nu}\partial_{\nu}
\Phi))\sqrt{\bb}}\right]  
\nonumber \\
+ 
 \sqrt{-g}\frac{
e^{-2\Phi}(4-\partial_{\mu}\Phi g^{\mu\nu}
\partial_{\nu}\Phi)}
{(1+\frac{1}{2}T^2(4-\partial_{\mu}\Phi g^{\mu\nu}
\partial_{\nu}\Phi))^2}+\nonumber \\
- \partial_{\mu}
\left[\frac{\sqrt{-g} T^2e^{-2\Phi}
g^{\mu\nu}\partial_{\nu}\Phi}
{(1+\frac{1}{2}T^2(4-\partial_{\mu}\Phi g^{\mu\nu}
\partial_{\nu}\Phi))^2}\right]
 \ . \nonumber \\
\end{eqnarray}
As we have shown  in
\cite{Kluson:2004ns} the equations of motions
(\ref{gequation}) and (\ref{phiequat}) have
following exact solution corresponding
to the linear dilaton
background 
\begin{equation}\label{linb}
\Phi=V_1x^1 \ , V_1=2 \ , g^{\mu\nu}=\eta^{\mu\nu} \ 
\end{equation}
together with the nonzero tachyon profile 
\begin{equation}\label{Tdil}
T=\mu e^{2 x^1} \ 
\end{equation}
that is also solution of the equation of motion
that arises from the action 
(\ref{Ttwo}) if we take into account the
ansatz (\ref{linb}) \cite{Kluson:2004ns}.
 According
to the standard dictionary we can interpret
(\ref{Tdil})  as the vacuum expectation
value of the tachyon field $T$
\begin{equation}\label{Tvac}
\left<T(x^1,t)\right>=\mu e^{2 x^1} \ .
\end{equation}
At this place we must stress one important
point considering the vacuum expectation value
of the tachyon field $T$ \cite{Ginsparg:is}.
 As was  argued
for example in  \cite{Natsuume:sp,Dhar:1997ze} 
the correct vacuum expectation
value of $T$, rather than given in  (\ref{Tvac}),
 should be equal to 
\footnote{Very nice arguments  based
on the analysis of  minisuperspace wave functions
in the $c=1$ models that support the
claim that  the vacuum expectation value of the tachyon
is equal 
$T\sim \phi e^{2\phi}$ for $\phi\rightarrow
-\infty$ were given in 
\cite{Seiberg:1990eb,Moore:1991ag,Seiberg:1992bj}.
On the other hand it was also mentioned in
\cite{Seiberg:1992bj} that the minisuperspace wave functions
satisfy the Wheeler-de-Witt equation
with the tachyon vacuum value $<T>=\mu e^{2\phi}$
rather then with $\left<T\right>=\mu \phi e^{2\phi}$.
It is not clear how this is consistent with 
the fact that the cosmological constant operator
in $c=1$ model should be $\phi e^{2\phi}$.}
\begin{equation}\label{Tper}
T(x^1)=4g_s^{-1}(2x^1+c)e^{2x^1} \ ,
c=1+\Gamma'(1)+\ln g_s \ ,
g^{-1}_s=\frac{|\mu|}{\sqrt{2\pi}} \ ,
\end{equation}
where $g_s$ is string coupling constant and
$\mu$ is level of Fermi sea.  
From the point of view of the tachyon effective
action (\ref{Ttwo}) one
can understand the discrepancy between
(\ref{Tvac}) and (\ref{Tper}) as follows. 
The calculations given in \cite{Natsuume:sp,Dhar:1997ze} 
were performed in the semi classical
limit  $g_s\ll 1$. Looking at (\ref{Tper}) we 
get  that 
  $\mu\gg 1$ and consequently the coefficient
in front of the linear term in (\ref{Tper})
is very large. 
To compare this result with the  effective
field theory  analysis we must mention that
the tachyon profile $T=(a+b\phi)e^{2\phi}$
is also solution of the equation of motion of
the   tachyon
effective action (\ref{Ttwo})  when  
 the background dilaton and metric are given
in (\ref{linb}). However we have shown in
 \cite{Kluson:2004ns} that 
such a  tachyon profile induces  large backreaction
on metric and dilaton and thus cannot be considered
as an exact solution of the action $S=S_{g,\Phi}+S_T$. 
This fact has simple explanation.  The proposal
for the 
tachyon effective action (\ref{Ttwo}) was based on previous
works considering tachyon effective actions on
unstable D-branes 
\cite{Kutasov:2003er,Niarchos:2004rw,Kluson:2004qy}.
It was argued in \cite{Kutasov:2003er,Niarchos:2004rw} that
 tachyon effective actions that have the tachyon
marginal profile as their exact solution correctly
capture tachyon dynamics  very close to the marginal
tachyon solution in the form $T\sim e^{\beta_{\mu}x^{\mu}}$.
In 2D bosonic string theory, $\beta_1=2$ and for the tachyon 
solution  $T=(a+b\phi)e^{2\phi}$ 
the condition that this profile is close to $T\sim e^{2\p}$
 implies that $b\ll 1$. Comparing with
the (\ref{Tper}) we see that this corresponds 
to $g_s\gg 1$ which is certainly out of region
of validity of the  classical analysis.
This result also implies that the tachyon effective
action (\ref{Ttwo}) cannot be derived from
the collective field theory that
arises in the classical description of the matrix
model \cite{Polchinski:uq,Das:1990ka}.
We must stress that remarks given above do not
solve the problems how to relate (\ref{Tvac}) and
(\ref{Tper}). In order to do that we should 
find how the tachyon effective action (\ref{Ttwo}) 
can be derived (We hope that this could be done.)
from the microscopic formulation of 2D string theory.
However in spite of these puzzling facts  
we still believe that the  action (\ref{Ttwo}) could  
 capture some problems in 2D string theory
that cannot be analysed in semi classical approximation.

To support this claim we begin to study the action
(\ref{Ttwo}) in more details. Let us introduce the scalar field $S(x)$ 
through 
\begin{equation}
T(\phi,t)=e^{2\phi}S(\phi,t) \ . 
\end{equation}
In terms of this field the tachyon effective action
(\ref{Ttwo})
in the linear dilaton background (\ref{linb})
takes simple form
\begin{equation}\label{Tthree}
S=-\int dt d\phi\left[
\sqrt{1-(\partial_t S)^2+(\partial_\phi S)^2}
-1\right] \ . 
\end{equation}
Now the  equation of motion that arises from (\ref{Tthree}) is
\begin{equation}\label{Tthreeeq}
\partial_{\mu}\left[\frac{\eta^{\mu\nu}
\partial_{\nu}S}{\sqrt{1+\eta^{\mu\nu}
\partial_{\mu}S\partial_{\nu}S}}\right]=0 \ . 
\end{equation}
The natural solution of this equation of motion is
\begin{equation}\label{Sc}
S_c=\mu=const. \ .
\end{equation}
 Moreover, in terms of the
field $S$ the stress energy tensor and dilaton
source have the form
\begin{equation}\label{TS}
T_{\mu\nu}=\left(-g_{\mu\nu}+2S^2V_{\mu}V_{\nu}+
\partial_{\mu}S\partial_{\nu}S-S^2V_{\nu}V_{\mu}\right)\sqrt{\bb}
+g_{\mu\nu}-S^2V_{\mu}V_{\nu}
\end{equation}
and
\begin{eqnarray}\label{DS}
J_{\Phi}=2\partial_{\mu}\left[S^2
g^{\mu\nu}V_{\nu}\sqrt{\bb}\right]
+g^{\mu\nu}
\partial_{\mu}S\partial_{\nu}S\frac{1}{\sqrt{\bb}}
-\nonumber \\
-\partial_{\mu}\left[g^{\mu\nu}
(V_{\nu}S+\partial_{\nu}S)\frac{1}{\sqrt{\bb}}\right]
-\partial_{\mu}\left[S^2g^{\mu\nu}V_{\nu}
\right]\ ,  \nonumber \\
\end{eqnarray}
where $\bb=1+\eta^{\mu\nu}
\partial_{\mu}S\partial_{\nu}S$.
It is easy to see that  components of
the stress energy tensor (\ref{TS}) 
and the dilaton source (\ref{DS}) are
equal to zero for $S_c=\mu$.  

It is interesting to study the action for fluctuation
modes around the classical solution $S_c$. If
we write the general field $S$ as
\begin{equation} 
S(\phi,t)=S_c+s(\phi,t) 
\end{equation}
and insert it to the action (\ref{Tthree}) we obtain
 the action for $s(\phi,t)$ that is exactly the same as the original
one. This can be regarded as a  consequence of the fact
that all  values of parameter  $\mu$ are equivalent.

In the end of this section we will briefly 
analyse another solution of the equation of motion
 (\ref{Tthreeeq}) that has the form  
\begin{equation}\label{Slin}
S=a\phi+b \ .
\end{equation}
 However
for such a tachyon profile the components of
the stress energy tensor are equal to
\begin{eqnarray}
T_{tt}=0 \ , T_{t\p}=0 \ , \nonumber \\
T_{\p\p}=1-\sqrt{1+a^2}+4(a\p+b)^2
(\sqrt{1+a^2}-1) \ . \nonumber \\
\end{eqnarray}
We see that $T_{\p\p}$ does not vanish. Then 
 one can expect strong backreaction on the 
metric and hence
the equation of motion of two dimensional gravity
is not satisfied for $g_{\mu\nu}=
\eta_{\mu\nu}$. One
can perform the same analysis for
the dilaton source with the same conclusion. 
For that reason we will not consider the 
solution (\ref{Slin}) further in this paper
since its meaning in the context of the proposed
model of 2D string theory is questionable.

In the next section we begin the analysis
of the quantum properties of the
action (\ref{Tthree}). We start with
the simple example of the quantum mechanics
of the zero mode.

\section{Quantum mechanics of zero mode}\label{third}
We start the analysis of the quantum properties
of the action (\ref{Tthree}) with the study of
the dynamics of the zero mode $S_0$. In this approximation
 we neglect
the dependence of the field $S$ on the spatial
coordinate $\p$. This
approach resembles minisuperspace analysis well
known from the study of two dimensional string
theory \cite{Seiberg:1990eb} and which
was recently used for analysis of the particle
production on S-branes \cite{Strominger:2002pc,Maloney:2003ck}.

To begin with let us consider 
the   Lagrangian for the zero mode $S_0$ 
in the form 
\begin{equation}
L=-\sqrt{1-\dot{S}_0^2}+1 \ .
\end{equation}
Then  the Hamiltonian is
equal to
\begin{equation}\label{hamzeromode}
H=P\dot{S}_0-L=\sqrt{1+P^2}-1 \ , 
\end{equation}
where we have used the fact that
the conjugate momentum $P$ is 
\begin{equation}
P=\frac{\dot{S}_0}{\sqrt{1-\dot{S}_0}} \ .
\end{equation}
The quantisation of the system given above is straightforward.
 The basic commutation
relation is
\begin{equation}
[\hat{S}_0,\hat{P}]=i \ 
\end{equation}
which implies that in the coordinate representation
$\hat{P}$ has the standard form
$P=-i\frac{\partial}{\partial S_0}$ when acts on the
wave function $\Psi(S_0)=\left<S_0|\Psi\right>$, where
$\ket{S_0}$ is eigenvector of the operator $\hat{S}_0$:
\begin{equation}
\hat{S}_0\ket{S_0}=S_0\ket{S_0} \ , 
\left<S_0|S_0'\right>=\delta(S_0-S_0') \ .
\end{equation}
From the  analysis of the classical equation
of motion  we know
that  the  tachyon configuration that
does not induce any backreaction on
the metric and dilaton corresponds to 
the Hamiltonian that is equal to zero.
Then it is natural to define 
the  vacuum state as an  eigenstate of the
Hamiltonian  
\begin{equation}
\hat{H}\Psi(S_0)=E\Psi(S_0)
\end{equation}
that has zero energy.
Now  (\ref{hamzeromode})   implies  
\begin{equation}
\hat{P}^2\ket{\Psi_0}=0\Rightarrow
-\frac{\partial^2\Psi_0(S_0)}{\partial^2S_0}=0
\end{equation}
that has general solution
\begin{equation}
\Psi_0(S_0)=\frac{1}{N}(aS_0+b) \ , 
\end{equation}
where $a \ , b$ are some constants and where 
  $N$ is normalisation factor. To clarify its
meaning note that the coordinate $S_0$
 takes values in the interval $(-\infty,\infty)$.
  Then the
normalisation condition implies
\begin{equation}
1=\int dS_0 |\Psi_0|^2
\Rightarrow N^2=\frac{2a^2}{3}L^3+2b^2L   \ ,
\end{equation}
where we have introduced the IR cut-off $L$.
We see that the wave function $\Psi_0(S_0)$ is
not normalisable in the limit $L\rightarrow \infty$. On
the other hand the expectation value of the
operator  $\hat{S}_0$ in the state $\ket{\Psi_0}$ is finite and 
it is
equal to
\begin{eqnarray}
\left<\Psi_0|\hat{S}_0|\Psi_0\right>=
\int dS_0 \Psi_0(S_0)S_0\Psi_0(S_0)=
\frac{4abL^3}{3N^2}=\nonumber \\
=
\frac{2b}{a(1+\frac{3b^2}{a^2L^2})}
\Rightarrow \frac{2b}{a}\equiv \mu \ , \mathrm{for}
\ 
L\rightarrow \infty \ . \nonumber \\ 
\end{eqnarray}
We must say few words  to the 
fact that  $\Psi_0(S_0)$ is
not normalisable. Usually such functions
are discarded from the Hilbert space
of given theory. On the other hand we know
from the analysis of two dimensional string
theory  that non normalisable states 
are  important for description of given theory. 
More precisely, 
it was argued in \cite{Seiberg:1990eb,Seiberg:1992bj}
 that since these wave functions
do not fluctuate they parametrise different
backgrounds of given  theory. 
As we will see more clearly in the
next section  the same conclusion can be
said about the vacuum states given above. 

Let us now consider following operator  
\begin{equation}\label{Os}
\hat{O}_{\epsilon}=e^{-i\epsilon\hat{P}} \ ,
\end{equation}
where $\epsilon$ is real parameter. 
Now we are going to argue that  this operator 
has  similar properties as the
spectral flow operator that was discussed in
the context of two dimensional string theory 
in \cite{DeWolfe:2003qf}.
First of all, it is easy to prove following commutation relation
\begin{equation}\label{SO}
[\hat{S}_0,\hat{O}_{\epsilon}]=\epsilon\hat{O}_{\epsilon} \ .
\end{equation}
Then we can  consider the  state
\begin{equation}
\ket{\Psi'_0}=\hat{O}_{\epsilon}\ket{\Psi_0} \ .
\end{equation}
Using (\ref{SO}) it is easy to 
 calculate  the  expectation value of
the operator $\hS_0$ in$\ket{\Psi'}$ 
\begin{eqnarray}\label{psiSn}
\left<\Psi'_0|\hat{S}_0|\Psi_0'\right>=
\mu+\epsilon \ .
\nonumber \\
\end{eqnarray}
Using also the fact that $\hat{O}_{\epsilon}$
commutes with $\hat{H}$ it easy to see that 
$\ket{\Psi'}$ is eigenstate of the Hamiltonian
with zero energy and consequently it can be
regarded as the new vacuum state characterised 
by the expectation value (\ref{psiSn}). In other
words, if we choose one particular vacuum state $\ket{\Psi_0}$ with the
expectation value $\left<\hS_0\right>=\mu$ one
can construct  family of all vacuum states
through the action of the operator $\hat{O}_{\epsilon}$
on $\ket{\Psi_0}$. We see
that all these states are equivalent and differ
in the expectation value of $\hS_0$
and consequently can be considered
as quantum analogues  of the classical solutions
(\ref{Sc}).

For consistency of the zero mode quantum
mechanics we   should show
that the vacuum states  do not induce
 any backreaction on metric and dilaton.
In order to see this we 
must calculate  the
vacuum expectation value of the zero mode truncation
of the stress energy tensor and the dilaton source. As follows from (\ref{TS})
the zero mode truncation of the stress energy tensor   
is equal to 
\begin{equation}
\hat{T}_{\mu\nu}=
\left(-g_{\mu\nu}+2\hat{S}_0^2V_{\mu}V_{\nu}
-\hat{S}_0^2V_{\nu}V_{\mu}\right)\sqrt{
1+\hat{P}_0^2}+g_{\mu\nu}-\hat{S}^2_0V_{\mu}V_{\nu}  
\ .
\end{equation}
For the spatial linear dilaton $V_{\phi}=2$ 
we get
\begin{eqnarray}\label{zerotensor}
\hat{T}_{tt}=\sqrt{1+\hat{P}^2}-1 \ ,
\nonumber \\
\hat{T}_{\phi\phi}=(-1+4\hat{S}_0^2)\sqrt{1+\hat{P}^2}
+1-4\hat{S}_0^2 \ , \nonumber \\
\hat{T}_{t\phi}=0 \ .  \nonumber \\
\end{eqnarray}
Since  $\hat{T}_{tt}$ is the same
as the Hamiltonian (\ref{hamzeromode})  we immediately
obtain that its vacuum
expectation value is equal to zero.
 For $\hat{T}_{\phi\phi}$ the action
of the square root on $\Psi_0$ is equal
to $\Psi_0$ and hence the vacuum expectation
value of $\hat{T}_{\phi\phi}$ is equal to zero as
well. This result implies that there is not
any backreaction of the vacuum states on the metric
even at the quantum mechanical level.

Now we are going to calculate the vacuum
expectation value of the zero mode truncation of the
 dilaton source (\ref{DS}) 
that is equal to
\begin{equation}\label{dilszero}
\hat{J}_{\Phi}=-\frac{\hat{P}^2}{(1+\hat{P}^2)^{3/2}}
+\frac{d}{dt}\left[\hat{S}_0\frac{\hat{P}}{
1+\hat{P}^2}\right] \ .
\end{equation}
We immediately see that
 the action of the first term on $\Psi_0$
 is equal to zero. Slightly
more complicated  is the calculation of the expectation
value  of the
second term in (\ref{dilszero}). It is clear that passing from
classical to quantum expression there is
potential ambiguity in ordering of $S$ and
$P$. We will adopt the ordering that
the operator $\hat{P}$ is always in the right
to the operator $\hat{S}_0$. In any case since the 
expectation value of the time derivative of any
operator $\hat{A}$  is equal to
\begin{equation}
\bra{\Psi}\frac{d\hat{A}}{dt}\ket{\Psi}=
\bra{\Psi}\left(\frac{\partial \hat{A}}{\partial t}
+i[\hat{H},\hat{A}]\right)\ket{\Psi}
\end{equation}
we  see that the vacuum expectation
value of the  second term in
(\ref{dilszero}) vanish as well due to the fact
that $\hat{H}\ket{\Psi_0}=0$ and the fact that
the operator $\hat{S}_0\frac{\hat{P}}{
1+\hat{P}^2}$ does not  depend on time. 

In order to obtain more precise picture
of the  vacuum states  
of  (\ref{Tthree}) we should  take 
into account the dynamics of the nonzero modes
as well. This  will be done in  the next
section.  
\section{Quantum mechanics of the tachyon
effective action}\label{fourth}
We begin  this section
with the construction of the Hamiltonian from Lagrangian
given in 
(\ref{Tthree})
\begin{eqnarray}\label{Hq}
H=\int d\p\left[\Pi\partial_t S-L\right]=  \nonumber \\
=\int d\phi\left[\sqrt{(1+(\partial_{\phi}S)^2)
(1+\Pi^2)}-1\right] \  , \nonumber \\
\end{eqnarray} 
where $\Pi(\p,t)=\frac{\delta \mathcal{L}}
{\delta \partial_t S(\p,t)}$ is  momentum conjugate to $S(\phi,t)$.
It turns out that it is convenient 
to  split the fields $S , \Pi$ into their zero mode parts and 
the fluctuation parts as 
\begin{eqnarray}
S(\p,t)=S_0(t)+\psi(\p,t) \ , 
S_0(t)=\frac{1}{V_\p}
\int d\p S(\p,t) \ , 
\int d\p \psi(\p,t)=0 \ , 
\nonumber \\
\Pi(\phi,t)=\frac{P(t)}{V_\p}+\pi(\phi,t) \ ,
P=\int d\p \Pi(\p,t) \ , \int d\p \pi(\p,t)=0 \ ,
\nonumber \\
\end{eqnarray}
where $V_\p$ is regularized volume of 
spatial section. 
In  quantum theory we 
regard the fields $S, \Pi$ as quantum 
operators $\hS(\phi,t) \ , \hP(\p,t)$ with 
the standard
equal time commutation
relations  
\begin{equation}
[\hS(\phi,t),\hS(\p',t)]=
[\hP(\p,t),\hP(\p',t)]=0
 \ ,[\hS(\p,t),\hP(\p',t)]=
i\delta(\p-\p') \  .
\end{equation}
Then we immediately get
\begin{equation}
[\hS_0,\hat{P}]=
\frac{1}{V_\p}\int d\p d\p'
[\hS(\phi),\hP(\p')]=\frac{i}{V_\p}
\int d\p d\p'
\delta(\p-\p')=i \ 
\end{equation}
and also
\begin{equation}\label{czerof}
[\hat{\psi}(\phi),\hat{P}]=
[\hS_0,\hat{\pi}(\p)]=0 \ .
\end{equation}
As we have argued in previous sections 
it is believed  that the tachyon
effective action (\ref{Ttwo}) is valid  for
description of  almost homogeneous 
field $S$. Then 
it is natural to restrict  ourselves in the Hamiltonian
(\ref{Hq})  to the
leading order terms in
$\partial_\p\psi, \pi$ . With this approximation
the quantum  Hamiltonian operator has the form
\begin{eqnarray}\label{hamquan}
\hat{H}=V_\p\left(\sqrt{1+\frac{\hat{P}^2}{V_\p^2}}-1\right)
+\frac{1}{2}\int d\p \left(
\frac{\hat{\pi}^2+(\partial_\p\hat{\psi})^2}
{\sqrt{1+\frac{\hat{P}^2}{V_\p^2}}}
+\frac{(\partial_\p\hat{\psi})^2\hat{P}^2}{V_\p^2
\sqrt{1+\frac{\hat{P}^2}{V_\p^2}}}\right) \ .\nonumber \\
\end{eqnarray}
Note that thanks to the commutation relations
(\ref{czerof}) and the fact that the Hamiltonian
does not contain zero mode operator $\hS_0$
there are not any problems with the
ordering of various operators in (\ref{hamquan}). 

In order to find ground states of the Hamiltonian
(\ref{hamquan}) we will work in the 
 Schr\"{o}dinger representation of the quantum field
theory. In this description the explicit time dependence
of the system is expressed through the time-dependent
wave functionals $\Psi(\phi,t)$. More precisely,
the time dependence of the wave functional is
determined by Schr\"{o}dinger equation  
\begin{equation}\label{Seq}
i\frac{\partial \Psi(\p,t)}{\partial t}=
\hat{H}(-i\frac{\delta }{\delta S_0} \ ,
S_0,-i\frac{\delta }{\delta \psi} ,
\psi)\Psi(\p,t) \ ,
\end{equation}
where $\hat{H}$ is given in (\ref{hamquan}).
Let us now  presume that the vacuum 
 wave functional has the 
form 
\begin{equation}\label{vacfun}
\Psi(S_0,\psi)=\Psi_0(S_0)\Phi(\psi) \ , 
\end{equation}
where the zero mode part $\Psi_0(S_0)$ obeys
\begin{equation}\label{zerop}
\sqrt{1+\frac{\hat{P}^2}{V_\p^2}}\Psi_0(S_0)=\Psi_0(S_0)
\ .
\end{equation}
From (\ref{zerop}) we see  that the vacuum
state  $\Psi_0(S_0)$ is solution of 
the differential equation
\begin{equation}
\hat{P}^2\Psi_0(S_0)=-\frac{\partial^2\Psi_0(S_0)}
{\partial S_0^2}=0 
\end{equation}
that is the same as the equation disused in the context
of the quantum mechanics of the zero mode
 in the previous section. 
Consequently the 
vacuum wave functional  $\Psi(S_0,\psi)$
is labelled with
the parameter $\mu$ that is vacuum
expectation value of the zero mode 
operator $\hS_0$. 
It is also clear that we can define the operator
$\hat{O}_{\epsilon}=e^{-i\epsilon \hat{P}}$ that
maps the vacuum state with $\left<\hS_0\right>=
\mu$ to the state 
$\Psi(\phi,t)'=\hat{O}_{\epsilon}\Psi_0(S_0)\Phi(\p,t)$.
Since 
$\hat{P}$ commutes with $\hat{\psi} \ ,
\hat{\pi}$, $\hat{O}_{\epsilon}$ 
acts on the zero mode part 
of the vacuum wave functional
only  we immediately see that $\hat{O}_{\epsilon}$
is the same as the operator (\ref{Os}).

In order to  determine the vacuum wave functional for
nonzero modes we use the Schr\"{o}dinger 
equation  (\ref{Seq}) that
for (\ref{vacfun}) is equal to   
\begin{equation}
\Psi_0(S_0)i\frac{\partial \Phi(\psi)}{\partial t}=
\Psi_0(S_0) \frac{1}{2}\int d\p\left(
\hat{\pi}^2+(\partial_{\p}\hat{\psi})^2\right)\Phi(\psi) \ .
\end{equation}
In the previous equation we have used the
fact that
\begin{eqnarray}
\hat{H}\Psi(S_0,\psi)=
\frac{1}{2}\int d\p\left(
\frac{\hat{\pi}^2+(\partial_\p\hat{\psi})^2}
{\sqrt{1+\frac{\hat{P}^2}{V_\p^2}}}
+\frac{(\partial_\p\hat{\psi})^2\hat{P}^2}{V_\p^2
\sqrt{1+\frac{\hat{P}^2}{V_\p^2}}}\right)\Psi_0(S_0)\Phi(\psi)=
\nonumber \\
=\Psi_0(S_0)\frac{1}{2}\int d\p\left(
\hat{\pi}^2+(\partial_{\p}\hat{\psi})^2\right)\Phi(\psi)
 \  \nonumber \\
\end{eqnarray}
which is a consequence of   (\ref{zerop}). 
Hence $\Phi(\psi)$ obeys the
equation
\begin{eqnarray}\label{vacflu}
i\frac{\partial \Phi(\psi,t)}{\partial t}=
\frac{1}{2}\int d\p\left(-\frac{\delta^2}{\delta^2\psi}+
(\partial_\p\psi)^2\right)\Phi(\psi) \Rightarrow
\nonumber \\
\Rightarrow
i\frac{\partial \Phi(\psi)}{\partial t}=\frac{1}{2}\int \frac{dk}{2\pi}
\left(-\frac{\delta^2}{\delta \alpha(k)
\delta \alpha(-k)}+k^2\alpha(k)\alpha(-k)\right)
\Phi(\alpha(k)) \ ,
\nonumber \\
\end{eqnarray}
where the second line in (\ref{vacflu})
is transformation of the first line to the 
momentum representation. 
The solution of (\ref{vacflu}) that corresponds
to the ground state is well known  
\cite{Guven:1987bx,Long:1996wf}. It can 
be shown that the vacuum wave functional
is equal to
\begin{equation}
\Phi(\psi)=N(t)\exp\left[-\frac{1}{2}\int \frac{dk}{(2\pi)}
\alpha(k)A(k)\alpha(-k)\right)\equiv N(t)E(\alpha(k)) \ ,
\end{equation}
where
\begin{eqnarray}
N(t)=C\exp\left[-i \frac{1}{2}\int 
\frac{dk}{(2\pi)} A(k)\right] \ ,  A(k)=|k| \ . \nonumber \\
\end{eqnarray}
In other words  the vacuum state for fluctuation
is ordinary vacuum wave functional for free, massless particles
in two dimensions. The time-dependent factor $N(t)$ 
can be written as 
$N(t)=e^{-iE_0t}$ where the vacuum energy of oscillators
is $E_0=\frac{1}{2}\int dk |k|=\frac{\Lambda^2}{2}$.
The vacuum energy depends on the
the cut-off $\Lambda$ that expresses the fact that the
tachyon effective
action (\ref{Tthree})
 is valid for tachyon fields close to the marginal
tachyon profile $e^{2\p}$. For fluctuation modes
this condition implies that    $\Lambda \ll 1$.

Let us review results obtained in this section.
We have got the family
of vacuum states of the tachyon effective action that are
labelled with the continuous parameter $\mu$.
We have argued  that $\mu$ corresponds
to the vacuum expectation value of the zero mode
operator $\hS_0$. We were also able to find
operator $\hat{O}_{\epsilon}$ that maps any 
vacuum state labelled with $\mu$ to another
vacuum state labelled with the parameter
$\mu+\epsilon$. We mean that existence
of this operators shows that 
 all vacuum states with different
$\left<\hS_0\right>$ are equivalent.
We can also alternatively  
interpret this result as the fact that
vacuum states with different $\mu$'s 
parametrise   superlescion sectors
of given theory. 
 We have also seen that 
the vacuum state for fluctuations is
ordinary  vacuum wave functional for massless scalar
field in Minkowski spacetime  with the
exception that there is a natural cut-off $\Lambda$ that
arises from the  limited
region of validity of the tachyon effective
 action (\ref{Tthree}).  Note that 
 the vacuum wave functional for fluctuations 
does not depend on $\mu$. 
\subsection{Calculation of the expectation values}
In the previous section we have found family of
vacuum states of the tachyon effective action
in the linear dilaton background. However as in
section (\ref{third})   we must show 
that these states do not induce backreaction on metric
and dilaton.  For that reason we will now calculate 
  the expectation
values of the dilaton source and stress energy tensor 
in  the state (\ref{vacfun}). 

To begin with  we  split the field $S$ into its  zero mode  and 
nonzero mode parts   and insert them into (\ref{TS}) 
so that for $V_\p=2$ we get classical components
of the stress energy tensor 
\begin{eqnarray}
T_{tt}(\p,t)=(1+\dot{S}^2_0+2(\partial_t\psi)\dot{S}_0
+(\partial_t\psi)^2)\sqrt{\bb}-1 \ , \nonumber \\
T_{\p\p}(\p,t)=(-1+4S_0^2+8S_0\psi+4\psi^2
-\partial_\p\psi\partial_\p\psi)\sqrt{\bb}+1
-4S_0^2-8S_0\psi-4\psi^2) \ , \nonumber \\
T_{t\p}(\p,t)=T_{\p t}(\p,t)=
\partial_\p\psi(\partial_tS_0+\partial_t\psi)
\sqrt{\bb} \ . \nonumber \\
\end{eqnarray}
As in previous section we will presume that
$\partial_\p\psi \ , \pi$  are small so that $\bb$ 
is equal to
\begin{equation}
\sqrt{\bb}=\frac{1}{\sqrt{1+\frac{P^2}{V_\p^2}}}
\left(1+\frac{1}{2}\left[(\partial_\p\psi)^2-\frac{\pi^2}
{1+\frac{P^2}{V^2_\p}}\right]\right) \ . 
\end{equation}
Then the quantum stress energy tensor
can be written as
\begin{eqnarray}\label{Tquan}
\hat{T}_{tt}=(1+(\partial_t\hat{S}_0)^2+2\partial_t\hat{\psi}
\partial_t\hat{S}_0
+(\partial_t\hat{\psi})^2)
\frac{1}{\sqrt{1+\frac{\hat{P}^2}{V^2_\p}}}
\left(1+\frac{1}{2}\left[(\partial_\p\hat{\psi})^2
-\frac{\hat{\pi}^2}
{1+\frac{\hat{P}^2}{V^2_\p}}\right]\right)
-1 \ , \nonumber \\
\hat{T}_{\p\p}=(-1+4\hat{S}_0^2+8\hat{S}_0\hat{\psi}
+4\hat{\psi}^2
-(\partial_\p\hat{\psi})^2)
\frac{1}{\sqrt{1+\frac{\hat{P}^2}{V_\p^2}}}
\left(1+\frac{1}{2}\left[(\partial_\p\hat{\psi})^2-\frac{\hat{\pi}^2}
{1+\frac{\hat{P}^2}{V_\p^2}}\right]\right)
+\nonumber \\
+1
-4\hat{S}_0^2-8\hat{S}_0\hat{\psi}-4\hat{\psi}^2 \ , \nonumber \\
\hat{T}_{t\p}=\hat{T}_{\p t}=
\partial_\p\hat{\psi}(\partial_t\hat{S}_0+\partial_t\hat{\psi})
\frac{1}{\sqrt{1+\frac{\hat{P}^2}{V_\p^2}}}
\left(1+\frac{1}{2}\left[(\partial_\p\hat{\psi})^2-\frac{\hat{\pi}^2}
{1+\frac{\hat{P}^2}{V_\p^2}}\right]\right)
 \ . \nonumber \\
\end{eqnarray}
We would like to stress that for zero mode
operators we adopt the same ordering
of various  operators as was defined in section
(\ref{third}) that means that
 $\hat{P}$ is always on the right with
respect to
$\hS_0$. On the other hand we will
argue that for nonzero modes the more
natural is to define the stress energy tensor and
the dilaton source as normal ordered quantities.  

Now we can proceed  to the  calculation of  the expectation value
of the components of the stress energy tensor
(\ref{Tquan}) in the vacuum state (\ref{vacfun}).
As the first step we  calculate the expectation value
with respect to the zero mode part $\Psi_0(S_0)$. 
Since $\hat{P}^2\Psi_0(S_0)=0$ we easily obtain
\begin{equation}\label{Bpsi}
\hat{\sqrt{\bb}}\Psi(S_0,\psi)=
\Psi_0(S_0)\left[1+
\frac{1}{2}((\partial_\p\hat{\psi})^2-\hat{\pi}^2)\right]\Phi(\psi) \ .
\end{equation}
Now let us consider the time derivative of 
$\hS_0$ that is equal to 
\begin{equation}\label{S0t}
\frac{d\hat{S}_0}{dt}=i[\hat{H},\hat{S}_0] \ .
\end{equation}
The vacuum expectation value of this
expression   is zero since
the vacuum state $\Psi_0$ is 
eigenstate of the Hamiltonian.  
On the other hand we can  calculate the right hand
side of the equation (\ref{S0t}) directly   
and we obtain that the commutator given there  is proportional
to $\hat{P}$. This result implies that
$(\partial_t\hS_0)^2\sim \hat{P}^2$ and consequently
its vacuum expectation value is equal to zero as well. 
Then the 
calculation of the vacuum expectation value of the zero mode 
operator is straightforward. When  we also restrict
ourselves to the terms of quadratic order in
$\hat{\psi} \ , \hat{\pi}$ we 
  obtain that
the vacuum expectation value of the stress energy tensor 
equal to
\begin{eqnarray}\label{Tfluct}
\left<T_{tt}(\phi,t)\right>=\frac{1}{2}\left<\hat{\pi}^2+
(\partial_\p\hat{\psi})^2\right>_{fluct}
\nonumber \\
\left<T_{\p\p}(\p,t)\right>=
(-1+4\mu^2)\frac{1}{2}\left<(\partial_\p\hat{\psi})^2
-\hat{\pi}^2\right>_{fluct}
-\left<(\partial_\p\hat{\psi})^2\right>_{fluct} \ ,  \nonumber \\
T_{\p  t}(\p,t)=
\left<\partial_\p\hat{\psi}\hat{\pi}\right>_{fluct} \ ,  \nonumber \\
\end{eqnarray}
where $\left<\dots\right>_{fluct}$ means the vacuum expectation
value with respect to the vacuum state of the
 fluctuation modes. In the previous expressions
 we have also used the
fact that
\begin{equation}
\partial_t\hat{\psi}=i[\hat{H},\hat{\psi}]=
\frac{\hat{\pi}}{\sqrt{1+\frac{\hat{P^2}}{V_\p}}}
\end{equation}
and consequently
\begin{equation}
\left<\partial_t\hat{\psi}\right>=
\left<\hat{\pi}\right>_{fluct} \ .  
\end{equation}
Following \cite{Guven:1987bx}
it is now 
easy to calculate the vacuum expectation
value in (\ref{Tfluct}) and we get
\begin{eqnarray} 
\left<T_{tt}(\p)\right>_{fluct}=
\int \frac{dk}{2\pi}
\frac{dk}{2\pi}e^{i\p(k+k')}\frac{1}{2}\left<
-kk'\alpha(k)\alpha(k')-\frac{\delta^2}{\delta \alpha(k)
\alpha(k')}\right>_{fluct} \ ,
\nonumber \\
\left<T_{\p\p}(\p)\right>_{fluct}=
(-1+4\mu^2)\int \frac{dk}{2\pi}
\frac{dk}{2\pi}e^{i\p(k+k')}\frac{1}{2}\left<
-kk'\alpha(k)\alpha(k')+\frac{\delta^2}{\delta \alpha(k)
\alpha(k')}\right>_{fluct}-\nonumber \\
-\int \frac{dk}{2\pi}
\frac{dk}{2\pi}e^{i\p(k+k')}
\left<kk'\alpha(k)\alpha(k')\right>_{fluct} \ ,
\nonumber \\
\left<T_{\p t}(\p)\right>_{fluct}=
-\int \frac{dk}{2\pi}\frac{dk'}{2\pi}
e^{i\p(k+k')}\left<ik\alpha(k)\frac{\delta}
{\delta \alpha (k')}\right>_{fluct} \ .
\nonumber \\
\end{eqnarray}
Using again the results given in \cite{Guven:1987bx}
\begin{equation}
\left<\alpha(k)
\alpha(k')\right>_{fluct}
=2\pi\frac{\delta(k+k')}{2|k|} \ ,
\left<\frac{\delta^2}{\delta \alpha(k)
\alpha(k')}\right>_{fluct}
=-\frac{k^2}{2|k|}2\pi\delta(k+k') \
\end{equation}
we finally obtain  
\begin{eqnarray}\label{Tvacfin}
 \left<T_{tt}(\p)\right>_{fluct}=\frac{\Lambda^2}{2} \ ,
\left<T_{\p\p}(\p)\right>_{fluct}=
-\frac{1}{2}\Lambda^2 \ , 
\left<T_{\p t}(\p)\right>_{fluct}=0 \ .
\nonumber \\
\end{eqnarray}
We see that the stress energy tensor
is diagonal and  its components 
scale with the cut-off $\Lambda$.
This is the same result  as in
standard quantum field theory calculation however the interpretation is
different. In ordinary QFT there is not
restriction on $\Lambda$ so that when we
remove the cut-off by taking $\Lambda\rightarrow
\infty$ we obtain
divergent contributions. One possibility how  to avoid
such a behaviour is to define field theory  operators
as normal ordered with respect to the vacuum state
which means that  their vacuum
expectation value is zero. Even 
if the  cut-off $\Lambda$ in (\ref{Tvacfin})
has clear meaning since it express
  the finite region of validity of
the tachyon effective action (\ref{Ttwo})
it is natural to consider stress energy
tensor  as normal ordered.
Then the vacuum expectation value of
the normal ordered stress energy tensor
 is equal to zero. In context of 2D effective
field theory this result implies that the vacuum
states do not induce any backreaction on
metric. 

As a next step we will calculate the vacuum
expectation value of the dilaton source
\begin{eqnarray}\label{dilq}
\hat{J}_{\Phi}(\p)=4\partial_{\p}\left[(\hat{S}_0
+\hat{\psi})^2
\sqrt{\bb}\right]
-
((\partial_t\hat{S}_0)+\partial_t\hat{\psi})^2\frac{1}{\sqrt{\bb}}
-\nonumber \\
+(\partial_\p\hat{\psi})^2\frac{1}{\sqrt{\bb}}
-\partial_{\p}\left[(2(\hat{S}_0+\hat{\psi})+\partial_\p
\hat{\psi})\frac{1}{\sqrt{\bb}}\right]+
\nonumber \\
+\partial_t\left[\partial_t(\hat{S}_0+\hat{\psi})
\frac{1}{\sqrt{\bb}}\right]
-2\partial_{\p}\left[(\hat{S}_0+\hat{\psi})^2\right] \ . \nonumber \\
\end{eqnarray}
We can proceed in the same way as in the calculation of the
expectation value of the stress energy tensor. More precisely,
 we begin with the calculation of the expectation value of
the zero mode wave function $\Psi_0(S_0)$.
If we again
restrict to the terms in the leading order in
$\partial\hat{\psi}\ , \hat{\pi} \ $    we obtain 
\begin{eqnarray}
\left<\hat{J}_{\Phi}\right>=
8\left<\hat{\psi}\partial_\p\hat{\psi}\right>_{fluct}-
\left<(\partial_t\hat{\psi})^2\right>_{fluct} 
\nonumber \\
+\frac{1}{2}\left<(\partial_\p \hat{\psi})^2+\hat{\pi}^2
\right>_{fluct}
-\left<\partial_\p[2(\mu+\hat{\psi})+\partial_\p\hat{\psi}]
\right>_{fluct} \nonumber \\
+\left<\partial_t^2\hat{\psi}\right>_{fluct}
-2\left<\partial_\p(\mu+\hat{\psi})^2\right>_{fluct} \ . 
\nonumber \\
\end{eqnarray}
This expression can be calculated in 
the same way as in case of the evaluation of
the vacuum expectation values of the stress
energy tensor. It is then easy to show
that the vacuum expectation value of
$\hat{J}$ is equal to zero.

Let us outline results obtained in this section.
 By explicit
calculations of the vacuum expectation values
of the normal ordered components of the stress
energy tensor we have shown that they
are equal to  zero and consequently these vacuum
states do not induce any backreaction on
metric. The same conclusion holds in case of
dilaton source as well. We mean that these
results provide further support for the
consistency of vacuum states of the  tachyon effective action
(\ref{Ttwo}). 
\section{Conclusion}\label{fifth}
In this article we have studied the
vacuum states of the  model of two dimensional
tachyon effective field theory that was proposed
in \cite{Kluson:2004ns}. The main goal was
to find whether there exists the similarity
between the ground states in the tachyon
effective action and the vacuum states in
the microscopic formulation of 2D theory
in terms of free fermions. We mean that
looking for  this relation could be useful for the
study of the non perturbative aspects of the
2D string theory. In fact, we have proposed the 
model of the tachyon effective action 
that has family of ground states 
that do not induce any backreaction
on metric and dilaton. We have also found
that there exists operator $\hat{O}_{\epsilon}$
that maps one ground state with parameter
$\mu$ to another one with parameter $\mu+\epsilon$.
In other words all ground states with different
values of $\mu$ are equivalent.

We mean that it would be very interesting to
find similar structure in the microscopic
formulation of 2D theory since we know that
the tachyon field arises from the bosonisation
in the dual matrix model. In particular, it
would be nice to find  operators
in the quantum field theory of free fermions
that are related to the operators $\hS_0 \ , 
\hat{O}_{\epsilon}$ in the effective
field theory description.  
We hope to return to this problem in the near future.

There are   many problems and open questions
that deserve to be studied further. 
One of the most serious one  is to find how
the effective field theory  
emerge from the
matrix model without restriction to the semi classical
approximation. We believe  that the approach developed in
\cite{Dhar:1992hr,Dhar:jc,Sengupta:1990bt,Das:1991qb,
Dhar:1992rs,Mandal:1991tz} could be helpful in 
solving of  this problem. 
Another open problem  is 
to find effective field theory 
  models of two dimensional OA and OB theories.
We expect  that construction of 
 these  models could be straightforward
extension of the model given in this paper when we
include terms containing 
the Ramond-Ramond field $C$ into the effective action. It would be
certainly very interesting to see whether one can find
such effective actions that have 
 backgrounds given in
\cite{Kapustin:2003hi,Gukov:2003yp,Davis:2004xb,
Thompson:2003fz,Strominger:2003tm}
 as their  exact solutions.  
\\
\\
{\bf Acknowledgement}
This work was supported by the
Czech Ministry of Education under Contract No.
14310006.

\end{document}